# Analogue cosmological particle creation in an ultracold quantum fluid of light

Jeff Steinhauer[1,2], Murad Abuzarli[1], Tangui Aladjidi[1], Tom Bienaimé[1], Clara Piekarski[1], Wei Liu[1], Elisabeth Giacobino[1], Alberto Bramati[1], and Quentin Glorieux[1]

[1]*Laboratoire Kastler Brossel, Sorbonne Université, CNRS, ENS-PSL Research University, Collège de France, Paris 75005, France*

[2]*Department of Physics, Technion—Israel Institute of Technology, Technion City, Haifa 32000, Israel*

In inflationary cosmology, the rapid expansion of the early universe resulted in the spontaneous production of cosmological particles from vacuum fluctuations, observable today in the cosmic microwave background anisotropies. The analogue of cosmological particle creation in a quantum fluid could provide insight, but an observation has not yet been achieved. Here we report the spontaneous creation of analogue cosmological particles in the laboratory, using a quenched 3-dimensional quantum fluid of light. We observe acoustic peaks in the density power spectrum, in close quantitative agreement with the quantum-field theoretical prediction. We find that the long-wavelength particles provide a window to early times, and we apply this principle to the cosmic microwave background. This work introduces a new quantum fluid, as cold as an atomic Bose-Einstein condensate.

Cosmological particle creation is an intriguing form of pair production from vacuum fluctuations, due to the expansion or contraction of the universe [1-6]. In the inflationary model of cosmology [7-10], the cosmological particles spontaneously created by tearing apart quantum fluctuations during the rapid cosmic expansion have an observable signature today, in the anisotropy of the cosmic microwave background (CMB) [11]. Since 1990, several successive space missions [12-15] have improved the resolution of the CMB power spectrum measurements, revealing the presence of characteristic peaks. These peaks are well described by acoustic oscillations in the quantum fluid of photons and baryons in the early universe [16], and suggest that cosmic inflation is the correct theory [11]. Analogue systems can explore this hypothesis, since they allow for measurement over time, which is impossible for the real universe, for which there is only one time of observation. Since the acoustic peaks are independent of the microscopic description of the medium, various quantum fluids were proposed for the study of cosmological particle creation in analogue universes [17-23]. In a two-dimensional atomic Bose-Einstein condensate, the out-of-equilibrium evolution after an interaction quench was reported, but the signal-to-noise ratio did not allow a quantitative comparison with cosmological particle creation [24]. In a recent experiment, a rapid switch in the trapping field of two ions led to phonon pair creation and formation of spatial entanglement [25]. However, this last configuration is limited to 1 dimension, and is not a quantum fluid.

In this work, we introduce a novel 3-dimensional quantum fluid of light, as coherent as an atomic Bose-Einstein condensate, and we observe time-resolved analogue cosmological particle creation out of vacuum fluctuations. Our quantum fluid is a near-resonant laser pulse traversing a warm atomic vapor cell, as illustrated in Fig. 1(a). Within the vapor cell, the two-body repulsive interactions between photons are mediated by the atoms, due to Kerr nonlinearity induced by the atomic resonance [26]. The interactions are suddenly quenched to zero when the laser beam exits the vapor cell [27]. This configuration mimics an expanding universe, as illustrated in Fig. 1(d), since a rapid reduction of the interactions causes a sudden red shift of the energy spectrum [17-21]. We also observe the reverse process at the cell entrance, in which the interaction suddenly appears, mimicking a contracting universe. By measuring the static structure factor, which is the spatial noise power spectrum, we demonstrate that both processes produce pairs of analogue cosmological particles. Also, the precision of our experiment allows for the observation of interferences between these two sets of analogue cosmological particles.

Our approach relies on the analogy between light propagation in a Kerr nonlinear medium and the temporal dynamic of an atomic Bose-Einstein condensate. The effective time is $\tau = z/c$, where $z$ is the position in the direction of propagation, and $c$ is the speed of light. This effective time is equivalent to true time for the sake of quantum mechanical quasiparticle creation [27, 28]. With no approximation other than the usual paraxial and slowly-varying envelope approximations [29], we extend the standard monochromatic limit [29] and find that our fluid is described by the 3-dimensional Gross-Pitaevskii equation

$$i\hbar \frac{\partial \psi}{\partial \tau} = -\frac{\hbar^2}{2m}\nabla^2 \psi + U(\mathbf{r},\tau)\psi + g(\mathbf{r},\tau)|\psi|^2 \psi \quad (1)$$

where $|\psi|^2$ is the volume density of the photons, $m$ is their effective mass, $U$ is an external potential, and $g|\psi|^2$ is the mean-field interaction energy. The three spatial dimensions of $\nabla$ correspond to the transverse coordinates $(x, y)$ and to $z' = \gamma(v_g t - c\tau)$, which is a coordinate comoving at the group velocity $v_g$, and compressed by a factor $\gamma$ due to group velocity dispersion [30].

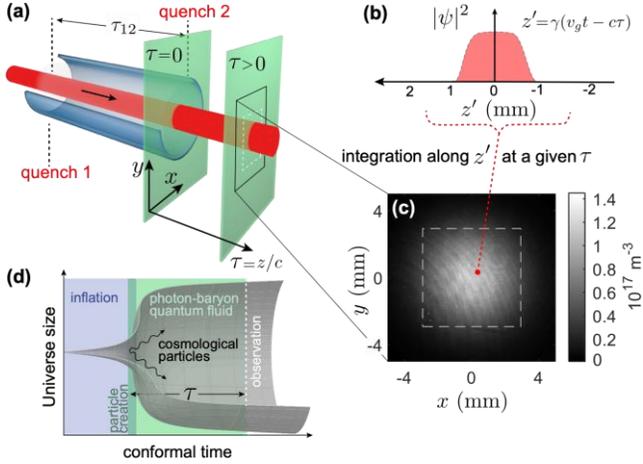

**Fig. 1. The analogue expansion. (a)** The fluid of light (red) is a laser pulse traversing a heated $^{85}$Rb vapor cell. The axial position gives the effective time $\tau$. The quenches occur at the entrance and exit of the vapor cell. $\tau = 0$ corresponds to quench 2. The time between the two quenches is $\tau_{12}$. **(b)** The true time gives an effective third spatial dimension $z'$. **(c)** Typical image of the fluid of light integrated along $z'$, given in units of photon density. An effective time $\tau = 103$ ps after quench 2 is shown. **(d)** Cosmological particle creation in the early universe. The gray surface indicates the radius of the universe.

To create the fluid of light, we use a 100 ns laser pulse with a 4 mm Gaussian waist and an intensity of 100 mW, propagating in an 85Rb vapor cell heated to 150 ºC. The laser is detuned -1.5 GHz (90 natural linewidths, 6 times the Doppler broadening) from the D2 resonance. The interaction energy is determined by the nonlinear change in the refractive index $\Delta n$, which is computed from the experimental parameters [30]. By taking into account the compression factor $\gamma$, this configuration leads to a weakly interacting photon gas with a thickness of 2 mm in the $z'$ coordinate, and a dimensionless interaction coefficient $\rho a_s^3 = 2\times10^{-12}$, where $\rho$ is the average photon density, and $a_s$ is the effective scattering length [30].

In each of the quenches, counter-propagating pairs of quasiparticles are produced in all three dimensions, in analogy with cosmological particle creation. We study the analogue cosmological particles using the static structure factor of a quantum fluid, in analogy with the CMB power spectrum. It is given by $S(k,\tau) = \langle |\delta\rho(k_x,k_y,k_{z'},\tau)|^2 \rangle / M$, where $\delta\rho(k_x,k_y,k_{z'},\tau)$ is the spatial Fourier transform of the density fluctuation at time $\tau$, and $M$ is the total number of particles in the fluid. With this definition, a zero-temperature, non-interacting gas has $S(k) = 1$, reflecting the presence of spatial shot noise. The operator $\hat{b}_\mathbf{k}^\dagger$ corresponds to the creation of a quasiparticle in mode $\mathbf{k} = (k_x, k_y, k_{z'})$ oscillating at frequency $\omega_k$. In the presence of quasiparticle populations $N \equiv \langle \hat{b}_\mathbf{k}^\dagger \hat{b}_\mathbf{k} \rangle$ and correlations $C \equiv \langle \hat{b}_\mathbf{k} \hat{b}_{-\mathbf{k}} \rangle$, the static structure factor within the Bogoliubov approximation is given by [30]

$$S(k) = 1 + 2N + 2\mathrm{Re}(Ce^{-i2\omega_k \tau}). \quad (2)$$

The populations and correlations are given by

$$N = \beta^2 + N_0(\alpha^2 + \beta^2) + 2\alpha\beta\,\mathrm{Re}(C_0) \quad (3)$$

$$C = \alpha\beta + C_0\alpha^2 + C_0^*\beta^2 + 2\alpha\beta N_0 \quad (4)$$

where $N_0$ and $C_0$ are the populations and correlations before the quench, respectively, and $\alpha$ and $\beta$ are the Bogoliubov coefficients [30]. For our series of two quenches, Eqs. 3 and 4 are applied twice. Since each quench either starts or ends with no interactions, α and β are the same Bogoliubov coefficients which diagonalize the Hamiltonian of a weakly-interacting quantum fluid [31]. In the absence of quasiparticles before a given quench, the pair production is spontaneous, and Eqs. 3 and 4 become $N = \beta^2$ and $C = \alpha\beta$. On the other hand, a distribution of quasiparticles before the quench, thermal or otherwise, will stimulate additional pairs.

The fluid of light is imaged on a sCMOS camera, as shown in Fig. 1(c). We tune the imaging system to pick out a certain $z$ after the cell, and the camera integrates over true time, as illustrated in Fig. 1(b). Thus, each image shows the density integrated in the $z'$ direction, at an effective time $\tau$ after the second quench. For each $\tau$, an ensemble of 200 images is obtained in 7 seconds, and the power spectrum $S(k_x, k_y, k_{z'} = 0)$ is computed by 2-dimensional Fourier transforms within the dashed square shown in Fig. 1(c). The computation partially removes the effects of any drifts such as thermal convection, and accounts for the measured quantum efficiency of the camera [30].

In Fig. 2(a) we observe ring patterns in $S(k_x, k_y, k_{z'} = 0)$, oscillating as a function of $k$. These oscillations are the experimental signature of analogue cosmological particle creation, in close analogy with the acoustic peaks in the angular spectrum of the CMB. They occur because the modes $k$ are generated synchronously at the moment of the quench, and oscillate with different frequencies $\omega_k$. The rings shrink with $\tau$ since lower frequencies take longer to develop oscillations. The radius of the first minimum is seen to be in good agreement with the theoretical prediction of Eq. 2, indicated by the dashed green curve. The azimuthal averages of the spherically-symmetric $S(k)$ are indicated in black in Fig. 2(b). The red curves are calculated from Eq. 2, taking into account the two quenches, and the variations in $\alpha$, $\beta$, and $\omega_k$ which result from the measured absorption [30].

The low-$k$ behavior of $S(k)$ provides a window into the early times before the quenches, since the frequency of these modes approaches zero, so the modes do not have sufficient time to evolve during the experiment. The cut-off frequency is on the order of $1/\tau$, which corresponds to the first peak in $S(k)$. Well below this $k$-value, Eq. 2 reduces to $S(k) = 1 + 2N_1$, where $N_1$ is the incoherent population before the first quench. Thus, the value of $S(k)$ gives a direct measure of $N_1$. Fig. 3(a) shows the $S(k)$ curves for all $\tau$ plotted together. We observe that $S(k)$ is at most 1.4 for low $k$, as indicated by the dashed green line, giving $N_1 \leq 0.2$. This value is finite, which implies a negligible thermal component, since a thermal population diverges like $1/k$. Furthermore, it is less than unity, implying that the spontaneous contribution dominates. Thus, the analogue cosmological particle creation is spontaneous in the first quench. This is verified by the blue and green curves in Fig. 3(c), which show that stimulation in the first quench by thermal noise and white noise, respectively, would produce larger values of $S(k)$ than those of the experiment, for low $k$.

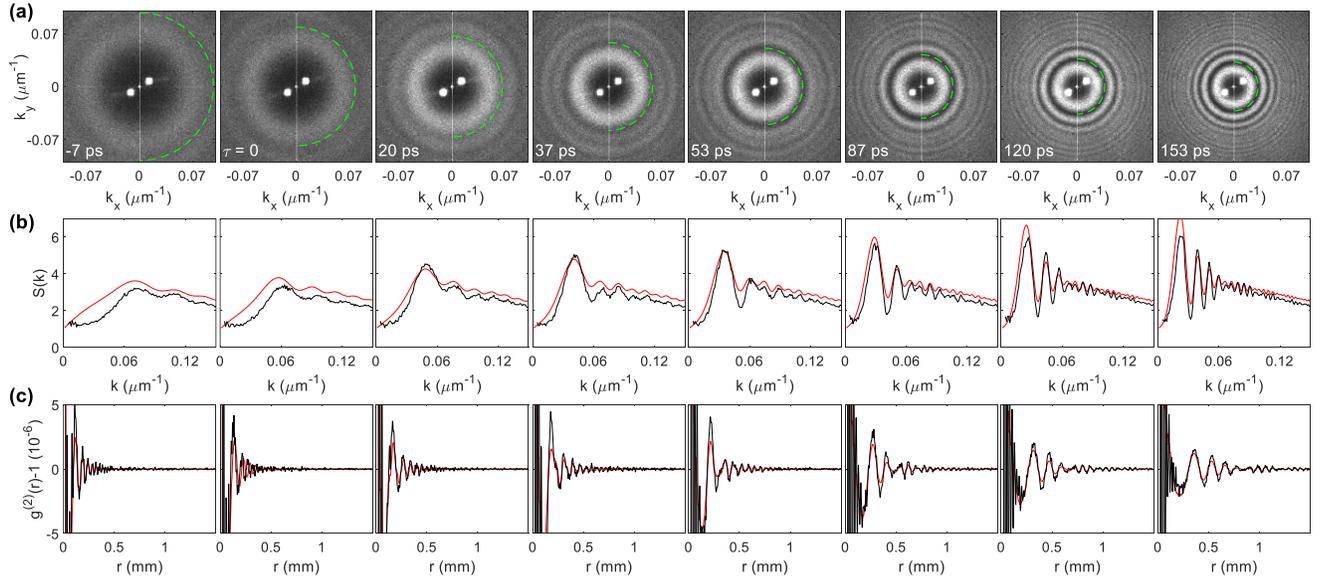

**Fig. 2. Analogue cosmological particle creation in a quantum fluid of light.** **(a)** The static structure factor $S(k_x, k_y, k_{z'} = 0)$ at various times after the second quench. The dashed green curve indicates the first minimum of the red curves in (b). The symmetric white points near the center of all panels are due to spurious fringes in the imaging system. **(b)** Radial profiles of (a). The black curves are the experimental data. The red curves are the prediction for analogue cosmological particle creation, from Eq. 2. **(c)** Density-density correlations. The experimental (black) and theoretical (red) curves are obtained from (b) by the spherical Fourier transform of Eq. 6.

The quasiparticles spontaneously created during the first quench stimulate pair creation in the second quench. However, if the particle production in the second quench were stimulated by the first-quench quasiparticles only, $S(k)$ would oscillate about unity, as indicated by the magenta curve in Fig. 3(d). The upward shift of $S(k)$ allows us to identify the presence of background quasiparticles, due to spontaneous and superradiant emission of photons from the atomic medium, which cause additional stimulation in the second quench. The downward slope of $S(k)$ observed at large $k$ is due to the finite resolution of the imaging system, measured to be 10 µm [30] and is included in all theoretical curves. Other than this slope, $S(k)$ oscillates about the value $1 + 2(N_1 + N_b)$, where $N_b$ is the background population present in the fluid between the two quenches. In our experimental configuration, we calculate that the contribution of spontaneous and superradiant emission leads to an incoherent population of $N_b$=1.3 [30]. The theoretical curves in Fig. 2(b) include this additional stimulation with no adjustable parameters, and confirm the origin of the background population. While this incoherent, flat spectrum of 1.3 quasiparticles per mode implies that the fluid is not in its ground state, like a finite-temperature Bose-Einstein condensate, it does not negate the oscillatory behavior of $S(k)$, and it even enhances the visibility of the oscillations. We can control this population by tuning the atomic density, the pulse duration, intensity, and detuning. In Fig. 3(e) we verify that this population vanishes for long weak pulses, as expected for spontaneous and superradiant emission.

Although our fluid of light is not in thermal equilibrium between the two quenches, we can put an upper limit on the effective temperature of the thermal component before the second quench. The blue curve in Fig. 3(d) includes thermal stimulation with an effective temperature $2mc_s^2 = 30$ mK, which results in a greatly enhanced first peak, absent from the experimental curve. Thus, we estimate the effective temperature of the thermal component to be less than $2mc_s^2$, as in an atomic Bose-Einstein condensate. For the second quench, the thermal fraction does not diverge like $1/k$ since the zero-temperature static structure factor in the fluid of light goes to zero for low $k$ [32].

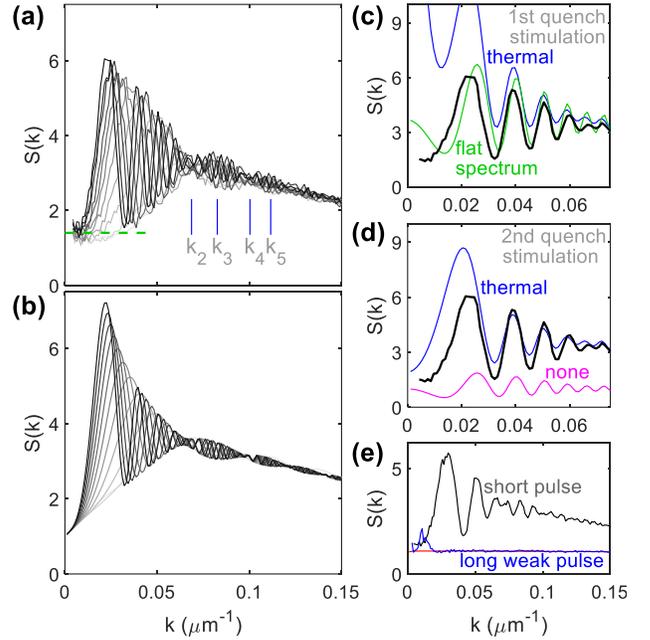

**Fig. 3. Spontaneous and stimulated cosmological particle production.** (a) The envelope of $S(k)$. The black curves of Fig. 2(b) are among the curves shown. Darker gray indicates later time. The low-$k$ limit is indicated by the green dashed line. The $k_p$ mark the nodes and antinodes. (b) The envelope of the theoretical curves. (c) The effect of stimulation in the first quench. The blue curve includes additional stimulation by a thermal distribution in the first quench. The green curve includes stimulation by a flat distribution in the first quench rather than the second. $\tau$ = 153 ps is shown. (d) The effect of stimulation in the second quench. The blue curve includes additional stimulation by a thermal distribution in the second quench. The magenta curve includes no extra stimulation in either quench. $\tau$ = 153 ps is shown. (e) Effect of the interactions. The black curve is from Fig. 2(b). The blue curve employs a pulse which is 500 times weaker and longer. The red curve is the theoretical prediction for the long, weak pulse. $\tau$ = 87 ps is shown.

Fig. 3(a) exhibits a beating pattern in the envelope of the various curves, resulting from interference between analogue cosmological particles created in the two quenches. The theoretical curves in Fig. 3(b) show a similar pattern. We can obtain a simple expression for the beating by neglecting absorption and approximating $\alpha \approx 1$ and $\beta \ll 1$, which is valid for all but the lowest values of $k$. This yields

$$S(k) = 1 + 2(N_1 + N_b) + 4\beta N_b \cos(2\omega_k \tau) + 4\beta(1 + 2N_1) \sin(\omega_{12_k} \tau_{12}) \sin(2\omega_k \tau + \omega_{12_k} \tau_{12}) \quad (5)$$

where $\omega_{12_k} = \sqrt{g\rho k^2/m + (\hbar k^2/2m)^2}$ and $\omega_k = \hbar k^2/2m$ are the Bogoliubov frequencies between the quenches and after the second quench, respectively, and $\tau_{12}$ is the time difference between the quenches. The last term in Eq. 5 results from the interference. The $\sin(\omega_{12_k} \tau_{12})$ factor is the envelope, which has nodes and antinodes at $\omega_{12_{k_p}} = \pi p/2\tau_{12}$, where $p$ is an integer. By identifying each $k_p$ as shown in Fig. 3(a), 4 points on the dispersion relation are found, as indicated by blue points in Fig. 4(a). These points agree well with the dispersion relation in the medium, calculated from the interactions, and indicated by the blue curve.

Fig. 4(b) shows the curves of Fig. 3(a), one above the other. By plotting the $S(k)$ values along the dashed line, we obtain the time dependence of a given mode $k$, as shown in Fig. 4(c). Each mode is seen to oscillate sinusoidally after the second quench. We observe as much as 3 full oscillation periods. The frequencies of the oscillations, indicated by the black curve in Fig. 4(a), agree well with the free-particle spectrum indicated by the magenta curve.

We also determine the spatial density correlations produced by the analogue cosmological particle creation. We derive the density-density correlation function $g^{(2)}(r)$ from $S(k)$ by the 3-dimensional spherically-symmetric Fourier transform

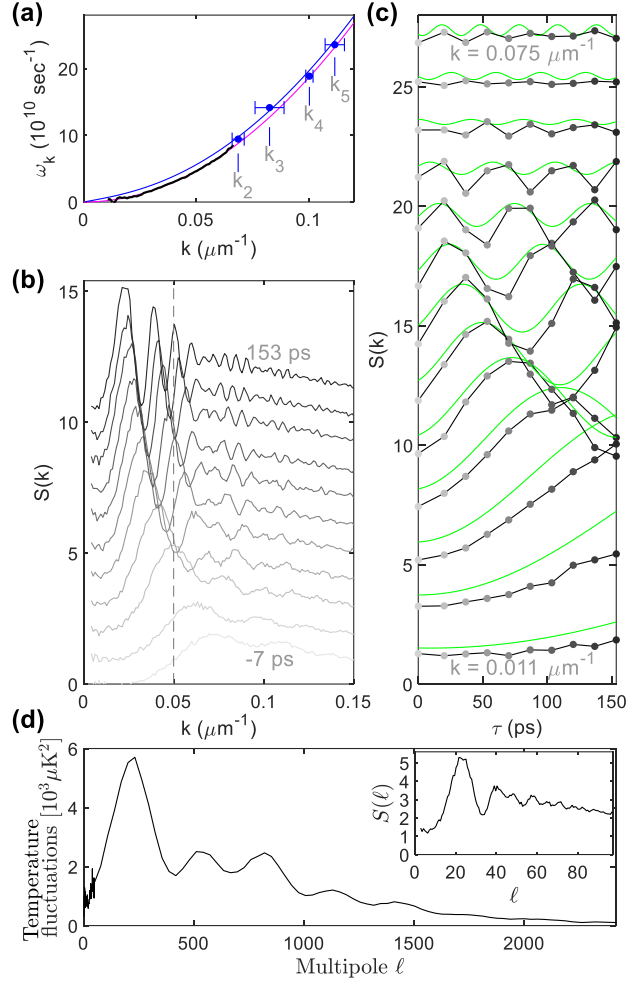

**Fig. 4. Individual modes of the analogue cosmological particles.** **(a)** The dispersion relation. The blue points are derived from the $k_p$ in Fig. 3(a). The black curve is obtained by sinusoidal fits to the curves in (c). The magenta curve is the free-particle dispersion relation. The blue curve is the dispersion relation in the interacting fluid of light. **(b)** The static structure factor at various times. The curves are from Fig. 3(a), and are shifted vertically. The vertical dashed line is used to find the values in (c). **(c)** Each curve shows the $\tau$-dependence of a definite $k$, given by the values along a vertical line in (b), such as the dashed line. The gray scale is the same as in (b). The $k$-values shown are equally spaced by $5.4\times10^{-3}$ μm$^{-1}$. The green curves are computed with Eq. 2. Each pair of black and green curves has been shifted vertically, with a spacing of 2. **(d)** The power spectrum of temperature fluctuations in the CMB, as a function of multipole number, from Ref. 15. The inset shows our measurement for $\tau = 53$ ps.

$$g^{(2)}(r) - 1 = \frac{1}{2\pi^2 \rho} \int dk \, k^2 \frac{\sin(kr)}{kr} [S(k) - 1]. \quad (6)$$

Figure 2(c) shows $g^{(2)}(r) - 1$, found by applying Eq. 6 to Fig. 2(b). The oscillations are spherical shells propagating outward. The correlations are seen to reach increasing distances as time increases. They are on the order of $10^{-6}$, which implies that the relative density fluctuations are on the order of $10^{-3}$. The oscillations are clear despite the small signal, due to the high sensitivity of the optical detection. The theoretical red curves are obtained by applying Eq. 6 to Eq. 2, and quantitative agreement with the experimental curves is seen.

Our experimental results open interesting prospects for cosmology. In particular, the window back in time discussed above also applies to the CMB angular spectrum from Ref. 15, shown in Fig. 4(d). The quench begins during the inflation period of the early universe, when the oscillation frequency $\omega_k$ of a given cosmological particle mode drops below the expansion rate of the universe, as illustrated in Fig. 1(d) [6]. It ends during the radiation-dominated period, when the expansion rate drops below $\omega_k$. The effective observation occurs a time $\tau$ later, when the photons decouple from matter [4]. The first peak near $\ell = 200$ in Fig. 4(d) corresponds to $\omega_k \tau = \pi$ [11]. The region well below the first peak is known to be a pristine imprint of the initial conditions just after the quench [15]. This region is similar to our measurement shown in the inset, plotted in terms of mode number $\ell \equiv kw/2\pi$, where $w$ is the waist of the laser beam (the radius of the analogue universe). This mode number is analogous to the multipole number in the CMB angular spectrum, when mapped back to the cosmological particle modes in the early universe [11]. We can use this region of the CMB spectrum to look even further back in time, since the low values in this region suggest vacuum before the quench, as in our measurement. Thus, the long-wavelength modes during inflation are likely near their vacuum state with a negligible thermal contribution, as predicted [4, 5].

In conclusion, we observe both spontaneous and stimulated analogue cosmological particle production in a quantum fluid of light. The particle production in the first quench is seen to be spontaneous, while the second includes stimulation by the first quench quasiparticles, as well as by an incoherent background. We quantitatively confirm the quantum field-theoretical prediction, with no free parameters. The long wavelength part of the spectrum provides a window into early times before the particle production, and we apply this principle to the CMB angular spectrum. This work establishes the paraxial fluid of light as a quantum fluid, and confirms that effective time is relevant for quantum field theory. The effective temperature is less than twice the chemical potential, which is comparable to many atomic Bose-Einstein condensates. On the other hand, the apparatus is an order of magnitude simpler, smaller, and less expensive. The direct detection of the photon fluid is also an advantage. Thus, the paraxial fluid of light presents opportunities for a wide class of experiments.

We thank I. Carusotto, T. Jacobson, M. Jacquet, and P.-É. Larré for helpful comments. This work received funding from the European Union Horizon 2020 research and innovation programme under grant agreement No 820392 (PhoQuS) and from the Region Île-de-France in the framework of DIM SIRTEQ. QG and AB thank the Institut Universitaire de France (IUF) for support.

1. E. Schrödinger, The proper vibrations of the expanding universe. *Physica* **6**, 899-912 (1939).
2. L. Parker, Particle creation in expanding universes. *Phys. Rev. Lett.* **21**, 562-564 (1968).
3. L. Parker, Particle creation in isotropic cosmologies. *Phys. Rev. Lett.* **28**, 705-708 (1972).
4. R. H. Brandenberger, Lectures on the theory of cosmological perturbations. *Lect. Notes Phys.* **646**, 127-167 (2004).
5. T. Jacobson, Introduction to quantum fields in curved spacetime and the Hawking effect, arXiv:gr-qc/0308048v3 (2004).
6. D. Campo and R. Parentani, Inflationary spectra and partially decohered distributions. *Phys. Rev. D* **72**, 045015 (2005).
7. A. A. Starobinskii, Spectrum of relic gravitational radiation and the early state of the universe. *JETP Lett.* **30**, 682-685 (1979).
8. K. Sato, Cosmological baryon-number domain structure and the first order phase transition of a vacuum. *Phys. Lett.* **99B**, 66-70 (1980).
9. A. H. Guth, Inflationary universe: A possible solution to the horizon and flatness problems. *Phys. Rev. D* **23**, 347-356 (1981).

10. A. D. Linde, A new inflationary universe scenario: A possible solution of the horizon, flatness, homogeneity, isotropy and primordial monopole problems. *Phys. Lett.* **108B**, 389-393 (1982).
11. W. Hu and M. White, Acoustic signatures in the cosmic microwave background. *Astrophys. J.* **471**, 30-51 (1996).
12. G. F. Smoot, et al., Structure in the COBE differential microwave radiometer first-year maps. *Astrophys. J.* **396**, L1-L5 (1992).
13. C. B. Netterfield, et al., A measurement by BOOMERANG of multiple peaks in the angular power spectrum of the cosmic microwave background. *Astrophys. J.* **571**, 604-614 (2002).
14. C. L. Bennett, et al., Nine-year Wilkinson microwave anisotropy probe (WMAP) observations: Final maps and results. *ApJS* **208**, 20 (2013).
15. N. Aghanim, et al., *Astronomy and Astrophysics* **641**, A1 (2019).
16. P. J. E. Peebles and J. T. Yu, Primeval adiabatic perturbation in an expanding universe. *Astrophys. J.* **162**, 815-836 (1970).
17. C. Barceló, S. Liberati, and M. Visser, Probing semiclassical analog gravity in Bose-Einstein condensates with widely tunable interactions. *Phys. Rev. A* **68**, 053613 (2003).
18. C. Barceló, S. Liberati, and M. Visser, Analogue models for FRW cosmologies. *Int. J. Mod. Phys. D* **12**, 1641-1649 (2003).
19. Petr. O. Fedichev and Uwe R. Fischer, "Cosmological" quasiparticle production in harmonically trapped superfluid gases. *Phys. Rev. A* **69**, 033602 (2004).
20. P. Jain, S. Weinfurtner, M. Visser, and C. W. Gardiner, Analog model of a Friedmann-Robertson-Walker universe in Bose-Einstein condensates: Application of the classical field method. *Phys. Rev. A* **76**, 033616 (2007).
21. R. Schutzhold and W. G. Unruh, *Analogue Gravity Phenomenology*, Ch. 3. Springer International Publishing, Switzerland (2013).
22. U. R. Fischer and R. Schutzhold, Quantum simulation of cosmic inflation in two-component Bose-Einstein condensates. *Phys. Rev. A* **70**, 063615 (2004).
23. A. Prain, S. Fagnocchi, and S. Liberati, Analogue cosmological particle creation: Quantum correlations in expanding Bose-Einstein condensates. *Phys. Rev. D* **82**, 105018 (2010).
24. C.-L. Hung, V. Gurarie, and C. Chin, From cosmology to cold atoms: Observation of Sakharov oscillations in a quenched atomic superfluid. *Science* **341**, 1213-1215 (2013).
25. M. Wittemer, F. Hakelberg, P. Kiefer, J.-P. Schröder, C. Fey, R. Schützhold, U. Warring, and T. Schaetz, Phonon pair creation by inflating quantum fluctuations in an ion trap. *Phys. Rev. Lett.* **123**, 180502 (2019).
26. Q. Fontaine, T. Bienaimé, S. Pigeon, E. Giacobino, A. Bramati, and Q. Glorieux, Observation of the Bogoliubov dispersion in a fluid of light. *Phys. Rev. Lett.* **121**, 183604 (2018).
27. P.-É. Larré and I. Carusotto, Propagation of a quantum fluid of light in a cavityless nonlinear optical medium: General theory and response to quantum quenches. *Phys. Rev. A* **92**, 043802 (2015).
28. Y. Lai and H. A. Haus, Quantum theory of solitons in optical fibers. I. Time-dependent Hartree approximation. *Phys. Rev. A* **40**, 844-853 (1989).
29. I. Carusotto, Superfluid light in bulk nonlinear media. *Proc. R. Soc. A* **470**, 20140320 (2014).
30. Supplementary material is given below.
31. N. Bogolubov, On the theory of superfluidity. *Journal of Physics* **11**, 23-32 (1947).
32. C. Piekarski, W. Liu, J. Steinhauer, E. Giacobino, A. Bramati, and Q. Glorieux, arXiv:2011.12935 (2020).
33. R. W. Boyd, *Nonlinear optics* (Elsevier, 2003).
34. L. Weller, R. J. Bettles, P. Siddons, C. S. Adams, and I. G. Hughes, Absolute absorption on the rubidium D1 line including resonant dipole–dipole interactions. *J. Phys. B: At. Mol. Opt. Phys.* **44**, 195006 (2011).
35. D. A. Steck, Rubidium 87 D line data (2003).
36. R. H. Dicke, Coherence in spontaneous radiation processes. *Phys. Rev.* **93**, 99-110 (1954).
37. P.-É. Larré, S. Biasi, F. Ramiro-Manzano, L. Pavesi, and I. Carusotto, Pump-and-probe optical transmission phase shift as a quantitative probe of the Bogoliubov dispersion relation in a nonlinear channel waveguide. *Eur. Phys. J. D* **71**, 146 (2017).

# Supplementary Material

3-dimensional Gross-Pitaevskii equation

It is known that the wave equation for a fluid of light takes the form of the following nonlinear Schrödinger equation, within the usual paraxial and slowly-varying envelope approximations [27, 29, 33]

$$i\frac{\partial \mathcal{E}}{\partial z} = -\frac{1}{2k_0}\nabla_\perp^2 \mathcal{E} + \frac{D_0}{2}\frac{\partial^2 \mathcal{E}}{\partial t^2} - \frac{i}{v_g}\frac{\partial \mathcal{E}}{\partial t} + U(\mathbf{r},t)\mathcal{E} + g(\mathbf{r},t)|\mathcal{E}|^2\mathcal{E} \tag{S1}$$

where $\mathcal{E}(x,y,z,t)$ is the electric field, $k_0$ is the wavenumber of the light, $\nabla_\perp \equiv (\partial_x, \partial_y)$, $D_0 = \partial^2 k_0/\partial \omega^2$ is the group velocity dispersion, $v_g = 0.007c$ is the group velocity, $U(\mathbf{r},t)$ is the external potential due to an applied index of refraction, and $g(\mathbf{r},t)$ is the effective coupling constant due to nonlinearity. The role of time is played by the spatial coordinate $z$ in the direction of propagation, and the fluid is 2-dimensional in the transverse coordinates $x$ and $y$. Equation S1 also contains first and second derivatives with respect to the true time $t$, but these are often neglected (the monochromatic approximation). However, the 3-dimensional nature of our study requires a different approach.

Firstly, we write Eq. S1 in energy units, as in the usual Gross-Pitaevskii equation,

$$i\hbar c\frac{\partial \psi}{\partial z} = -\frac{\hbar^2}{2m}\nabla_\perp^2 \psi - \frac{\hbar^2}{2m_{z'}v_g^2}\frac{\partial^2 \psi}{\partial t^2} - \frac{i\hbar c}{v_g}\frac{\partial \psi}{\partial t} + U(\mathbf{r},t)\psi + g(\mathbf{r},t)|\psi|^2\psi \tag{S2}$$

where $\psi(x,y,z,t)$ is the macroscopic wavefunction, normalized such that the mean of $|\psi|^2$ is the photon density $\rho \approx 7\times10^{16}$ m$^{-3}$. Also, the effective photon mass is $m = \hbar k_0/c = 2.8\times10^{-36}$ kg, and $m_{z'} = -\hbar/cv_g^2 D_0$. The mean-field interaction energy (the chemical potential) is given by $g(\mathbf{r},\tau)\rho = -\hbar c k_0 \Delta n/n$, and the external potential is given by $U(\mathbf{r},\tau) = -\hbar c k_0 \delta n/n$, where $\delta n$ is an applied change in the index of refraction. Furthermore, the healing length is given by $\xi = \hbar/mc_s = 60$ µm, where $c_s = \sqrt{g(\mathbf{r},\tau)\rho/m} = c\sqrt{-\Delta n/n}$ is the speed of sound for the Bogoliubov excitations. The length scale associated with group velocity dispersion is given by $\xi_{z'} = \xi(m/m_{z'})^{1/2} = 6$ mm (27). In order for Eq's. S1 and S2 to be valid, the frequency interval $v_g/2\pi\xi_{z'} = 50$ MHz should be much less than the 1.5 GHz detuning, which is indeed the case. The effective s-wave scattering length is $a_s = mg(\mathbf{r},\tau)/4\pi\hbar^2 = 3.1\times10^{-10}$ m.

Secondly, rather than neglecting the time derivatives, we make the coordinate transformation $(z,t) \to (z',\tau)$, where the effective time is $\tau = z/c$ ($z$ plays the role of time), and

$$z' \equiv \sqrt{\frac{m_{z'}}{m}}(v_g t - z) \tag{S3}$$

where the first factor is $\gamma$ in the main text. This coordinate $z'$ is comoving with the light at the group velocity, and is compressed due to group velocity dispersion. The relationship between $z$ and $z'$ is illustrated in Fig. S1. The square pulse in Fig. S1(a) is stationary and compressed when plotted versus $z'$, as shown in Fig. S1(b). The second factor in Eq. S3 is from Ref. 28. The transformation is

$$\begin{pmatrix} \frac{\partial \psi}{\partial z} \\ \frac{\partial \psi}{\partial t} \end{pmatrix} = \begin{pmatrix} \frac{\partial z'}{\partial z} & \frac{\partial \tau}{\partial z} \\ \frac{\partial z'}{\partial t} & \frac{\partial \tau}{\partial t} \end{pmatrix} \begin{pmatrix} \frac{\partial \psi}{\partial z'} \\ \frac{\partial \psi}{\partial \tau} \end{pmatrix} = \begin{pmatrix} -\sqrt{\frac{m_{z'}}{m}} & \frac{1}{c} \\ v_g\sqrt{\frac{m_{z'}}{m}} & 0 \end{pmatrix} \begin{pmatrix} \frac{\partial \psi}{\partial z'} \\ \frac{\partial \psi}{\partial \tau} \end{pmatrix} \tag{S4}$$

Applying the transformation twice gives

$$\frac{\partial^2 \psi}{\partial t^2} = v_g^2 \left(\frac{m_{z\prime}}{m}\right) \frac{\partial^2 \psi}{\partial z\prime^2} \tag{S5}$$

Inserting $\partial\psi/\partial z$, $\partial\psi/\partial t$, and $\partial^2\psi/\partial t^2$ from Eq's. S4 and S5 into Eq. S2 gives Eq. 1, the 3-dimensional Gross-Pitaevskii equation. All three space coordinates $x$, $y$, and $z'$ have the same mass $m$ and healing length $\xi$. Thus, we have recast Eq. S1 into the 3-dimensional Gross-Pitaevskii equation without any further approximation. The mapping carries on to the quantized field, as shown in Ref. 27.

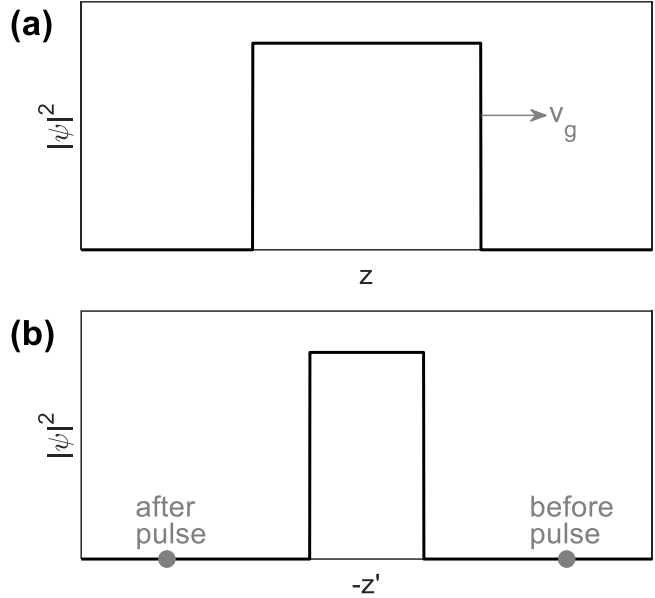

**Fig. S1. The comoving and compressed coordinate.** (a) A square pulse as a function of $z$. (b) The same pulse as a function of $z'$.

The determination of $\Delta n$

The interaction between photons is quantified by the nonlinear contribution to the index of refraction $n$, given by $\Delta n = n(I) - n(0)$, where $I$ is the laser intensity. We would like to express $\Delta n$ in terms of easily measurable quantities. The index of refraction is given by $n = \sqrt{1 + \text{Re}(\chi)}$, where $\chi$ is the atomic susceptibility. Since $n \approx 1$, one obtains $\Delta n = \Delta[\text{Re}(\chi)]/2$. Furthermore, the absorption coefficient is given by $\alpha_a = k_0 \text{Im}(\chi)$. Thus,

$$\Delta n = \frac{\alpha_a}{2k_0} \frac{\Delta[\text{Re}(\chi)]}{\text{Im}(\chi)}$$

Also, $\chi$ is proportional to $(i - 2\delta/\Gamma)\left[1 + (2\delta/\Gamma)^2 + I/I_{\text{sat}}\right]^{-1}$, where $\delta$ is the detuning from resonance, $\Gamma$ is the linewidth, and $I_{\text{sat}}$ is the saturation intensity. This gives

$$\Delta n = \frac{\alpha_a}{2k_0} \frac{-\frac{\frac{2\delta}{\Gamma}}{1+\left(\frac{2\delta}{\Gamma}\right)^2 + \frac{I}{I_{\text{sat}}}} + \frac{\frac{2\delta}{\Gamma}}{1+\left(\frac{2\delta}{\Gamma}\right)^2}}{\frac{1}{1+\left(\frac{2\delta}{\Gamma}\right)^2 + \frac{I}{I_{\text{sat}}}}} \tag{S6}$$

The detuning is δ = -1.5 GHz relative to the $^{85}$Rb cooling transition, and the self-broadened linewidth is Γ/2π = 16 MHz (34), for the vapor cell temperature of 150 ℃, corresponding to an atomic density of 1×10$^{20}$ m$^{-3}$. The intensity is given by $I = 2P/\pi w^2$, where the waist of the beam is $w$ = 4 mm, and the laser power $P$ decays exponentially due to absorption, from 100 mW at the entrance to the vapor cell, to 20 mW at the exit. We estimate $I_{\text{sat}}$ to be the far-detuned, π-polarized value, 25 Wm$^{-2}$ [35]. The absorption coefficient is given by $\alpha_a = -(\ln T)/L$, where $T$ = 0.2 is the transmission through the vapor cell of length $L$ = 10 mm. The wavenumber is given by $k_0 = 2\pi/\lambda$, where $\lambda$ = 780 nm. Equation S6 yields $\Delta n$ = -8.6×10$^{-6}$ and -1.7×10$^{-6}$ for the entrance and exits of the vapor cell, respectively. We have neglected the effect of optical pumping into the dark ground state. Via measurements of $\Delta n$ and $\alpha_a$, we find the optical pumping time to be a few microseconds, so optical pumping is negligible during the 100 ns pulse employed in this work.

Spontaneous and superradiant emission

The absorption of the laser pulse implies that photons were scattered by the atomic vapor. Early in the pulse, the number of previously scattered photons is small, so the emission is spontaneous. As the number of photons $N_m$ emitted into each single mode grows, the emission rate increases by a factor $1 + N_m$, due to superradiance [36].

The total number of photons $N_s$ emitted during the pulse is equal to the number of photons removed from the pulse by absorption,

$$N_s = \frac{P_L t_p}{\hbar c k_0}(1-T)$$

where $t_p$ is the length of the pulse, and $P_L = P \, \text{erf}^2(L_s/\sqrt{2}w)$ is the laser power within the dashed square in Fig. 1(c), with dimension $L_s$ = 6 mm. We are interested in the number of photons per mode $N_m$. The number $N_s$ is readily observable experimentally, and it essentially fixes $N_m$, regardless of whether the photons were emitted spontaneously or superradiantly. Nevertheless, we explicitly include superradiance in the calculation.

Only those photons with the same frequency and polarization as the fluid of light become quasiparticles. Thus, the absorption and emission must return the atom to the original hyperfine $F, m_F$ level. The ground state of $^{85}$Rb has $n_{\text{gs}}$=12 such levels. Furthermore, only $q$ = ½ of the emitted photons have the relevant polarization. Also, the width of a mode is given by $\delta k = 2\pi/L_s$, and the total number of modes is $n_m = 4\pi k_0^2/\delta k^2 = 4\pi L_s^2/\lambda^2$. The total number of photons emitted into each mode, with the correct frequency and polarization, is given by

$$N_t = N_s q/n_{\text{gs}} n_m = \frac{q\lambda^3 P t_p}{8\pi^2 \hbar c L_s^2 n_{\text{gs}}}(1-T)\,\text{erf}^2(L_s/\sqrt{2}w). \tag{S7}$$

However, the photons have a finite lifetime $\Delta t = 1/\Delta\omega$, where $\Delta\omega = 6\times 10^5$ sec$^{-1}$ is the spectral width of the laser. We can find the corresponding reduction in the number of photons per mode by considering the rate at which photons are created and lost, given by

$$\dot{N}_m = R(1 + N_m) - \frac{N_m}{\Delta t} \tag{S8}$$

where $R$ is the rate of spontaneous emission, and the first term includes superradiance. The rate is given by

$$R = t_p^{-1} \ln(1 + N_t) \tag{S9}$$

which ensures that the total number of photons produced by the first term in Eq. S8 is $N_t$. The solution of Eq. S8 at $t = t_p$ is

$$N_m = \frac{R}{\Delta\omega - R}\left[1 - e^{-t_p(\Delta\omega - R)}\right]. \tag{S10}$$

For the $t_p = 100$ ns pulses used in Fig. 2, Eq's. S7, S9, and S10 give $N_m = 1.3$. On the other hand, the blue curve in Fig. 3(d) employs pulses which are 500 times longer and weaker, corresponding to $N_m = 0.04$.

Computation of $S(k_x, k_y, k_{z\prime} = 0)$

The power spectrum (static structure factor) $S(k_x, k_y, k_{z\prime})$ of a system of $M$ particles (photons in our case) is given by

$$S(k_x, k_y, k_{z\prime}) = \frac{\langle |\delta\rho(k_x, k_y, k_{z\prime})|^2 \rangle}{M} \tag{S11}$$

where $\delta\rho(k_x, k_y, k_{z\prime}) = \int dx\, dy\, dz'\, \delta\rho(x, y, z')\, e^{-i(k_x x + k_y y + k_{z\prime} z')}$ and the density fluctuation $\delta\rho(x, y, z') = \rho(x, y, z') - \langle\rho(x, y, z')\rangle$. Setting $k_{z\prime} = 0$, one obtains the 2-dimensional Fourier transform

$$\delta\rho(k_x, k_y, k_{z\prime} = 0) = \int dx\, dy\, \delta\tilde{\rho}(x, y)\, e^{-i(k_x x + k_y y)} \tag{S12}$$

where $\tilde{\rho}(x, y) = \int dz'\, \rho(x, y, z')$ is the number density integrated in the $z'$ direction, a quantity we measure directly on the camera.

The density fluctuation $\delta\tilde{\rho} = \tilde{\rho} - \langle\tilde{\rho}\rangle_5$ is computed for each image, where $\langle\tilde{\rho}\rangle_5$ is the average of 5 adjacent images rather than the average $\langle\tilde{\rho}\rangle$ over the entire ensemble. This technique reduces the effects of drifts in the experimental parameters. As mentioned in relation to Fig. 2(c), the relative density fluctuation $\delta\tilde{\rho}/\langle\tilde{\rho}\rangle$ is on the order of $10^{-3}$, so small drifts can play a role. For example, thermal convection of the $^{85}$Rb gas may induce small changes in the shape of the fluid of light from image to image. The 2-dimensional Fourier transform of Eq. S12 is computed for each image within the dashed square of Fig. 1(c). The power spectrum $S(k_x, k_y, k_{z\prime} = 0)$ is computed by Eq. S11. The use of $\langle\tilde{\rho}\rangle_5$ rather than $\langle\tilde{\rho}\rangle$ reduces the fluctuations by a factor of 4/5. Thus, $S(k_x, k_y, k_{z\prime} = 0)$ is multiplied by 5/4 to correct this effect.

Furthermore, the finite quantum efficiency $Q = 0.485$ of the camera tends to randomize the photon density and bring $S(k_x, k_y, k_{z\prime} = 0)$ closer to unity. Thus, $S(k_x, k_y, k_{z\prime} = 0) - 1$ is multiplied by the factor $1/Q$.

The static structure factor as a measure of 2-mode fluctuations

We can gain additional insight into the power spectrum by noting that it measures the relative fluctuations between two regions of the fluid of light. The vector $(k_x, k_y, k_{z'} = 0)$ defines the two areas, indicated by red and green in Fig. S2. The power spectrum is essentially given by $S(k_x, k_y, k_{z'} = 0) = \langle (N_r - N_g)^2 \rangle / \langle N_r + N_g \rangle$, where $N_r$ and $N_g$ are the numbers of photons in the total red and green regions. The quantity $N_r - N_g$ could be computed by multiplying the image by a square wave and summing, but sine and cosine functions are used, giving a Fourier transform.

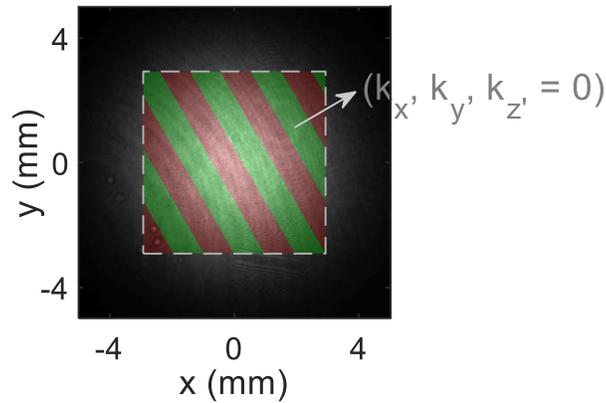

**Fig. S2. The power spectrum as a measure of relative fluctuations between two areas.** The total red and green regions of the image indicate the two areas defined by the vector $(k_x, k_y, k_{z'} = 0)$.

Measuring the imaging resolution

The resolution of the imaging system is determined from images of very small rubidium droplets on the window of the vapor cell, as shown in Fig. S3(a). The sizes of the droplets are below the resolution, so the images show the point spread function of the imaging system. The Fourier transform of the images gives the response of the imaging system as a function of $k$. The magnitude squared of the 2-dimensional Fourier transform is computed for each image, and the average over all images is found. The azimuthal average of this quantity is shown in Fig. S3(b). The dashed curve is a Gaussian fit, normalized to unity, which is taken as the response of the imaging system. The Gaussian is of the form $\exp(-R^2 k^2/4)$, where $R = 10$ µm is the measured resolution. Throughout this work, the theoretical curves for $S(k) - 1$ are multiplied by this Gaussian.

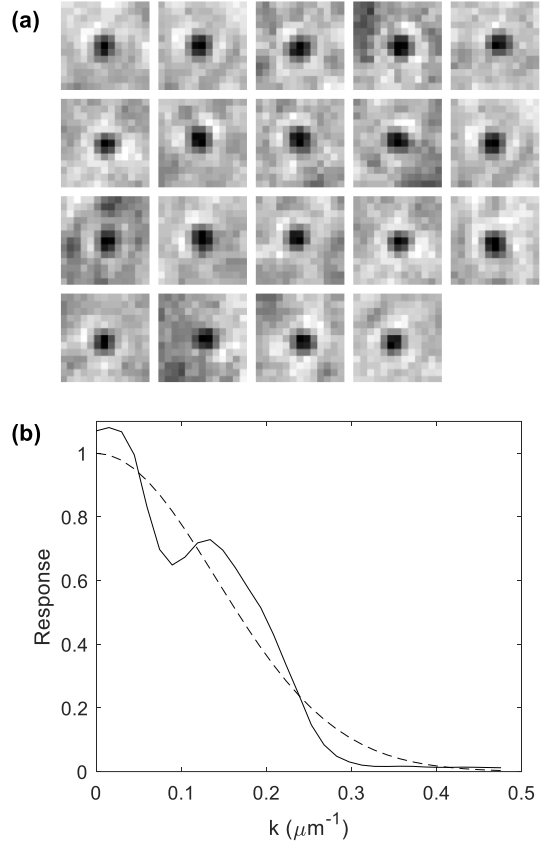

**Fig. S3. The resolution of the imaging system. (a)** Images of tiny rubidium droplets. Each image shows the point spread function of the imaging system. Each image has dimensions 85 µm × 85 µm. **(b)** The response of the imaging system. The solid curve is determined from the images in (a). The dashed curve is a Gaussian fit.

Derivation of Eq's. 3 and 4

The annihilation operators are given by $\hat{a}_\mathbf{k}$ (before the quench), and $\hat{b}_\mathbf{k}$ (after the quench). The population and correlations before the quench are $N_0 \equiv \langle \hat{a}_\mathbf{k}^\dagger \hat{a}_\mathbf{k} \rangle = \langle \hat{a}_{-\mathbf{k}}^\dagger \hat{a}_{-\mathbf{k}} \rangle$ and $C_0 \equiv \langle \hat{a}_\mathbf{k} \hat{a}_{-\mathbf{k}} \rangle = \langle \hat{a}_\mathbf{k}^\dagger \hat{a}_{-\mathbf{k}}^\dagger \rangle^*$, respectively. The operators before and after the quench are related by the Bogoliubov transformation $\hat{b}_\mathbf{k} = \alpha \hat{a}_\mathbf{k} + \beta \hat{a}_{-\mathbf{k}}^\dagger$, where the Bogoliubov coefficients are given by $\alpha = (\sqrt{S_0} + 1/\sqrt{S_0})/2$, and $\beta = (\sqrt{S_0} - 1/\sqrt{S_0})/2$, where $S_0 = (\xi k)^2/\sqrt{4(\xi k)^2 + (\xi k)^4}$, and $\xi$ is the healing length just after the first quench [31]. Since $\hat{b}_\mathbf{k}^\dagger = \alpha \hat{a}_\mathbf{k}^\dagger + \beta \hat{a}_{-\mathbf{k}}$, one obtains $\langle \hat{b}_\mathbf{k}^\dagger \hat{b}_\mathbf{k} \rangle = \alpha^2 \langle \hat{a}_\mathbf{k}^\dagger \hat{a}_\mathbf{k} \rangle + \beta^2 (1 + \langle \hat{a}_\mathbf{k}^\dagger \hat{a}_\mathbf{k} \rangle) + 2\alpha\beta \, \text{Re}\langle \hat{a}_\mathbf{k} \hat{a}_{-\mathbf{k}} \rangle$, which is Eq. 3. One also obtains $\langle \hat{b}_\mathbf{k} \hat{b}_{-\mathbf{k}} \rangle = \alpha^2 \langle \hat{a}_\mathbf{k} \hat{a}_{-\mathbf{k}} \rangle + \beta^2 \langle \hat{a}_\mathbf{k}^\dagger \hat{a}_{-\mathbf{k}}^\dagger \rangle + \alpha\beta(1 + 2\langle \hat{a}_\mathbf{k}^\dagger \hat{a}_\mathbf{k} \rangle)$, which is Eq. 4.

## $S(k)$ including absorption

The annihilation operators are given by $\hat{a}_\mathbf{k}$ (before the vapor cell), $\hat{b}_\mathbf{k}$ (in the vapor cell), and $\hat{c}_\mathbf{k}$ (after the vapor cell).

**Before the vapor cell (the initial state)**

The population is $N_1 \equiv \langle \hat{a}_\mathbf{k}^\dagger \hat{a}_\mathbf{k} \rangle = \langle \hat{a}_{-\mathbf{k}}^\dagger \hat{a}_{-\mathbf{k}} \rangle$, which may be a thermal distribution or otherwise. The population is assumed to be incoherent, so the correlations are $\langle \hat{a}_\mathbf{k} \hat{a}_{-\mathbf{k}} \rangle = 0$.

**First quench**

The first quench occurs at an effective time $\tau_0 = -L/c$. The operators before and after the quench are related by the Bogoliubov transformation $\hat{b}_\mathbf{k} = \alpha(\tau_0)\hat{a}_\mathbf{k} - \beta(\tau_0)\hat{a}_{-\mathbf{k}}^\dagger$, where the Bogoliubov coefficients are given by $\alpha(\tau_0) = (\sqrt{S_0} + 1/\sqrt{S_0})/2$, and $\beta(\tau_0) = (\sqrt{S_0} - 1/\sqrt{S_0})/2$, where $S_0 = (\xi k)^2/\sqrt{4(\xi k)^2 + (\xi k)^4}$, and $\xi$ is the healing length just after the first quench [31]. The population after the first quench is given by $\langle \hat{b}_\mathbf{k}^\dagger \hat{b}_\mathbf{k} \rangle = \alpha^2(\tau_0)N_1 + \beta^2(\tau_0)(1 + N_1) + N_b$, where the first two terms are from the Bogoliubov transformation, and $N_b$ is the background distribution of quasiparticles inside the vapor cell, thermal or otherwise. The correlations after the first quench are given by $\langle \hat{b}_\mathbf{k} \hat{b}_{-\mathbf{k}} \rangle = -\alpha(\tau_0)\beta(\tau_0)(1 + 2N_1)$. These populations and correlations are the initial state for the second quench.

**Second quench**

The second quench occurs at $\tau = 0$, and the operators before and after are related by the Bogoliubov transformation $\hat{c}_\mathbf{k} = \alpha(0)\hat{b}_\mathbf{k} + \beta(0)\hat{b}_{-\mathbf{k}}^\dagger$, where the Bogoliubov coefficients include the effects of propagation and absorption in the vapor cell. We find that the Bogoliubov coefficients obey the wave equation [37]

$$i\frac{\partial}{\partial \tau}\begin{bmatrix}\alpha(\tau) \\ \beta(\tau)\end{bmatrix} = \mathcal{K}(\tau)\begin{bmatrix}\alpha(\tau) \\ \beta(\tau)\end{bmatrix} \tag{S13}$$

where $\mathcal{K}(\tau) \equiv \begin{pmatrix} \frac{ic\alpha_a}{2} + \frac{\hbar k^2}{2m} + ck_0\Delta n & ck_0\Delta n \\ -ck_0\Delta n & \frac{ic\alpha_a}{2} - \frac{\hbar k^2}{2m} - ck_0\Delta n \end{pmatrix}$

We find

$$\begin{bmatrix}\alpha(\tau) \\ \beta(\tau)\end{bmatrix} = \prod_{q=M}^{0} e^{-i\mathcal{K}(\tau_q)\delta\tau} \begin{bmatrix}\alpha(\tau_0) \\ \beta(\tau_0)\end{bmatrix} e^{c\alpha\tau/2} \tag{S14}$$

where the first two factors are the solution of Eq. S13 from [37], the last factor is added to ensure that $|\alpha(\tau)|^2 - |\beta(\tau)|^2 = 1$, $\delta\tau = (\tau - \tau_0)/Q$, and $\tau_q = q\delta\tau$. The integer $Q$ should be large enough to ensure

convergence. Eq. S14 is evaluated at $\tau = 0$ to obtain the Bogoliubov coefficients for the second quench, $\alpha(0)$ and $\beta(0)$.

The populations after the second quench are given by $\langle \hat{c}_\mathbf{k}^\dagger \hat{c}_\mathbf{k} \rangle = |\alpha(0)|^2 \langle \hat{b}_\mathbf{k}^\dagger \hat{b}_\mathbf{k} \rangle + |\beta(0)|^2 (1 + \langle \hat{b}_\mathbf{k}^\dagger \hat{b}_\mathbf{k} \rangle) + 2 \langle \hat{b}_\mathbf{k} \hat{b}_{-\mathbf{k}} \rangle \operatorname{Re}[\alpha(0)\beta^*(0)]$, which becomes

$$\langle \hat{c}_\mathbf{k}^\dagger \hat{c}_\mathbf{k} \rangle = |\beta(0)|^2 + [|\alpha(0)|^2 + |\beta(0)|^2][\alpha^2(\tau_0) N_1 + \beta^2(\tau_0)(1 + N_1) + N_b]$$
$$- 2\alpha(\tau_0)\beta(\tau_0)(1 + 2N_1) \operatorname{Re}[\alpha(0)\beta^*(0)] \tag{S15}$$

The correlations just after the second quench are given by $\langle \hat{c}_\mathbf{k} \hat{c}_{-\mathbf{k}} \rangle = [\alpha^2(0) + \beta^2(0)]\langle \hat{b}_\mathbf{k} \hat{b}_{-\mathbf{k}} \rangle + \alpha(0)\beta(0)(1 + 2\langle \hat{b}_\mathbf{k}^\dagger \hat{b}_\mathbf{k} \rangle)$, which becomes

$$\langle \hat{c}_\mathbf{k} \hat{c}_{-\mathbf{k}} \rangle = -[\alpha^2(0) + \beta^2(0)]\alpha(\tau_0)\beta(\tau_0)(1 + 2N_1)$$
$$+ \alpha(0)\beta(0)\{1 + 2[\alpha^2(\tau_0) N_1 + \beta^2(\tau_0)(1 + N_1) + N_b]\} \tag{S16}$$

**Power spectrum**

In the Bogoliubov approximation, the Fourier transform of the density operator is given by $\rho_\mathbf{k} = \sqrt{N}[\alpha(\tau) + \beta(\tau)](\hat{c}_\mathbf{k}^\dagger + \hat{c}_{-\mathbf{k}})$. The power spectrum a time $\tau$ after the second quench is given by $S(\mathbf{k}) = \langle \rho_\mathbf{k} \rho_{-\mathbf{k}} \rangle / N = [\alpha(\tau) + \beta(\tau)]^2 [1 + 2\langle \hat{c}_\mathbf{k}^\dagger \hat{c}_\mathbf{k} \rangle + 2\operatorname{Re}(\langle \hat{c}_\mathbf{k} \hat{c}_{-\mathbf{k}} \rangle e^{-i2\omega_k \tau})]$, where the $\exp(-i\omega_k \tau)$ dependence of $\hat{c}_\mathbf{k}$ is included. Since there are no interactions after the second quench, $\alpha(\tau) = 1$ and $\beta(\tau) = 0$, and the power spectrum becomes

$$S(\mathbf{k}) = 1 + 2\langle \hat{c}_\mathbf{k}^\dagger \hat{c}_\mathbf{k} \rangle + 2\operatorname{Re}(\langle \hat{c}_\mathbf{k} \hat{c}_{-\mathbf{k}} \rangle e^{-i2\omega_k \tau})$$

where $\langle \hat{c}_\mathbf{k}^\dagger \hat{c}_\mathbf{k} \rangle$ and $\langle \hat{c}_\mathbf{k} \hat{c}_{-\mathbf{k}} \rangle$ are given by Eq's. S15 and S16, respectively.